\documentclass[runningheads]{llncs}
\usepackage[T1]{fontenc}
\usepackage{graphicx}
\usepackage{amsmath}
\newif\iffinal
\finaltrue
\newcommand{\cmtid}{60}
\begin{document}
%
\thispagestyle{empty}
{\noindent\Large Springer Copyright Notice}\\[1pt]

{\noindent Copyright (c) 2024 Springer

\noindent This work is subject to copyright. All rights are reserved by the Publisher, whether the whole or part of the material is concerned, specifically the rights of translation, reprinting, reuse of illustrations, recitation, broadcasting, reproduction on microfilms or in
any other physical way, and transmission or information storage and retrieval, electronic adaptation, computer software, or by similar or dissimilar methodology now known or hereafter developed.}\\[1em]

{\noindent\large Accepted to be published in: 2024 27th Iberoamerican Congress on Pattern Recognition (CIARP'24), Nov 26--29, 2024.}\\[1in]

{\noindent Cite as:}\\[1pt]

{\setlength{\fboxrule}{1pt}
 \fbox{\parbox{0.95\textwidth}{G. B. Carvalho and J. Almeida, ``Exploiting the Segment Anything Model (SAM) for Lung Segmentation in Chest X-ray Images,'' in \emph{2024 27th Iberoamerican Congress on Pattern Recognition (CIARP)}, Talca, Chile, 2024, pp. 1--14}}}\\[1in] 

{\noindent BibTeX:}\\[1pt]

{\setlength{\fboxrule}{1pt}
 \fbox{\parbox{1.1\textwidth}{
 @InProceedings\{CIARP\_2024\_Carvalho,

 \begin{tabular}{lll}
  & author    & = \{G. B. \{Carvalho\} and
               J. \{Almeida\}\},\\

  & title     & = \{Exploiting the Segment Anything Model (SAM) for \\
  &           & \ \ \ \ \ Lung Segmentation in Chest X-ray Images\},\\

  & pages     & = \{1--14\},\\

  & booktitle & = \{2024 27th Iberoamerican Congress on Pattern Recognition (CIARP)\},\\

  & address   & = \{Talca, Chile\},\\

  & month     & = \{November 26--29\},\\

  & year      & = \{2024\},\\

  & publisher & = \{\{Springer\}\},\\

  \end{tabular}

\}
 }}}

\clearpage

\title{Exploiting the \textit{Segment Anything Model}~(SAM) for Lung Segmentation in Chest X-ray Images}
\titlerunning{Exploiting the SAM for Lung Segmentation in Chest X-ray Images}
%
\iffinal

\author{
Gabriel Bellon de Carvalho \and
Jurandy Almeida
}
\authorrunning{G. Carvalho and J. Almeida}
%
\institute{Department of Computing\\
Federal University of São Carlos -- UFSCar\\
18052-780, Sorocaba, SP -- Brazil\\
\email{gabrielbellon@estudante.ufscar.br, jurandy.almeida@ufscar.br}\\}

\else

  \author{Paper ID: \cmtid \\ }

\fi
\maketitle              
\begin{abstract}

\textit{Segment Anything Model}~(SAM), a new AI model from Meta AI released in April 2023, is an ambitious tool designed to identify and separate individual objects within a given image through semantic interpretation. 
The advanced capabilities of SAM are the result of its training with millions of images and masks, and a few days after its release, several researchers began testing the model on medical images to evaluate its performance in this domain. 
With this perspective in focus -- i.e., optimizing work in the healthcare field -- this work proposes the use of this new technology to evaluate and study chest X-ray images.
The approach adopted for this work, with the aim of improving the model's performance for lung segmentation, involved a transfer learning process, specifically the fine-tuning technique. 
After applying this adjustment, a substantial improvement was observed in the evaluation metrics used to assess SAM's performance compared to the masks provided by the datasets.
The results obtained by the model after the adjustments were satisfactory and similar to cutting-edge neural networks, such as U-Net.

\keywords{Deep Learning \and Segment-Anything Model (SAM) \and Medical Image Analysis \and Lung Segmentation \and Chest X-Ray Images}
\end{abstract}

\section{Introduction}

\textit{Segment Anything Model}~(SAM)~\cite{sam} is a tool that, since its release in April 2023, has proven to be very promising in the task of image segmentation. Its approach involves using a variety of input prompts to identify different objects in images, such as points and bounding boxes. To predict the masks, SAM uses three components: (\textit{i}) an image encoder, (\textit{ii}) a prompt encoder, and (\textit{iii}) a mask decoder. Additionally, the model can automatically segment anything in an image and generate multiple valid masks for ambiguous inputs, which is innovative in the field.

Given this and the immense amount of data used in its training -- 11 million images and over 1 billion masks~\cite{sam} -- many researchers have recognized the potential of this technology in the medical field and have begun to investigate its effectiveness in this area. However, despite having this large volume of data in its training, there are no medical images among the domains in which SAM was trained, which makes its generalization ability moderate when it comes to this area~\cite{estudo2_sam,estudo1_sam}.

This study aims to advance the application of SAM in the field of medical image analysis, especially, for lung segmentation in chest X-ray images. 
Understanding the effectiveness of SAM in this domain is of paramount importance in the development of new technologies for the diagnosis, treatment and follow-up of lung diseases.
We finetune SAM on two collections of chest X-ray images, known as the Montgomery and Shenzhen datasets~\cite{dados}.
This well-established practice aims to leverage the benefits of representations previously learned on a larger database to optimize the training of a network on a smaller dataset.
Our exploration also involved testing SAM across such datasets using various input prompts, like bounding boxes and individual points.
The obtained results show that our finetuned SAM can perform similar to state-of-the-art approaches for lung segmentation, like U-Net~\cite{unet}.

\section{Related Work}

Several studies have conducted a comprehensive evaluation of SAM on a variety of medical image segmentation tasks~\cite{estudo2_sam,estudo1_sam}, demonstrating that the model achieved satisfactory segmentation results, especially on targets with well-defined boundaries. However, it is evident that SAM has certain limitations due to the lack of contour in the regions of the images in question, making it difficult to identify certain patterns such as the shapes of organs and tissues~\cite{estudo1_sam}.

Among such studies, it is worth mentioning the work of Ma~and~Wang~\cite{medsam}. They introduce \textit{MedSAM}, which was developed on an unprecedented set of over 1 million medical image-mask pairs. Also, they evaluated the fine-tuning of the model, the same technique adopted in this work for chest X-ray images, and the obtained results demonstrate the great potential of SAM in medicine. In most of the 86 tasks evaluated, \textit{MedSAM} ranked first, surpassing the performance of specialized models, like U-Net~\cite{unet} and DeepLabV3+~\cite{deeplab}.

In spited of all the advances, SAM achieved a maximum F1-Score of 60\% for lung segmentation of chest X-ray images using points as input, indicating poor performance, as highlighted in the work of He~et~al.~\cite{acuracia_sam}. This study also revealed that the model's performance with bounding box prompts was even worse.

\section{Segment Anything Model~(SAM)}

Referred to as SAM, the \textit{Segment Anything Model} was trained on a dataset containing over a billion masks~\cite{sam}, presenting itself as an innovative proposal in the field of segmentation. The primary purpose of this model is to generate a valid mask for any input image, relying on two essential functionalities.

\begin{figure}[!htb]
    \centering
    \begin{tabular}{cc}
        \includegraphics[width=0.48\textwidth]{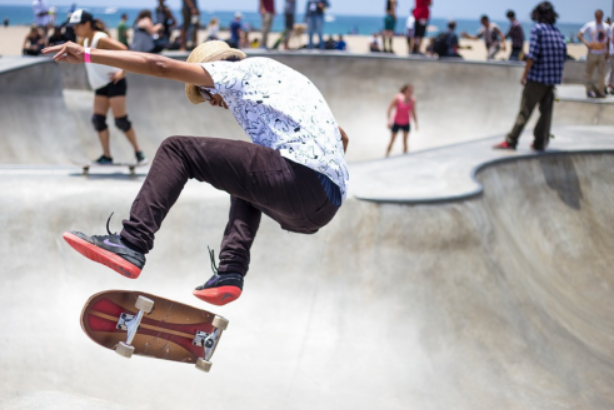} &
        \includegraphics[width=0.48\textwidth]{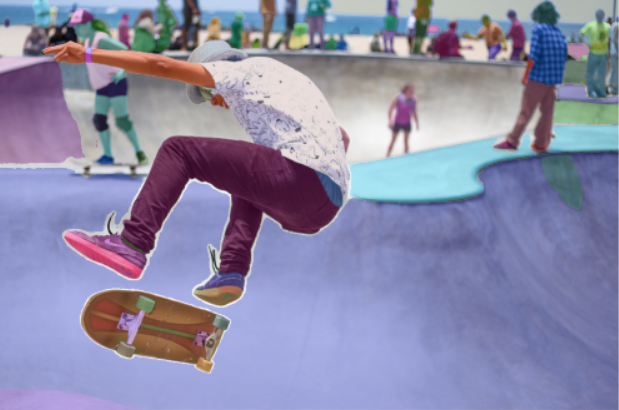} \\
        (a) original image & (b) panoptic segmentation \\
    \end{tabular}
    \caption{Example of a given image and its corresponding masks after a panoptic segmentation. Adapted from Google Images.}
    \label{fig:sam_cachorro}
\end{figure}

In the first of these, \textit{panoptic segmentation}, the goal is to automatically generate various masks for the objects identified in an image. The result of its execution is a list of masks, where each mask is a dictionary with descriptor fields. It is important to note that there are several adjustable parameters in the automatic mask generation, allowing control over interesting factors such as the density of sampling points and limits for removing duplicate masks. An example of using this functionality can be seen in Fig.~\ref{fig:sam_cachorro}.

\begin{figure}[!htb]
    \centering
    \begin{tabular}{cc}
        \includegraphics[width=0.48\textwidth]{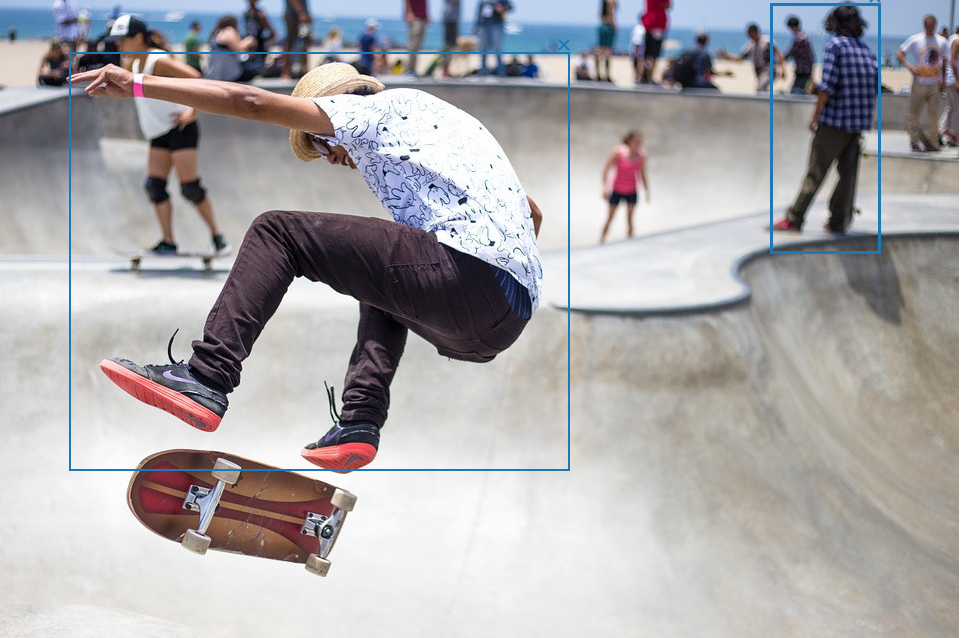} &
        \includegraphics[width=0.48\textwidth]{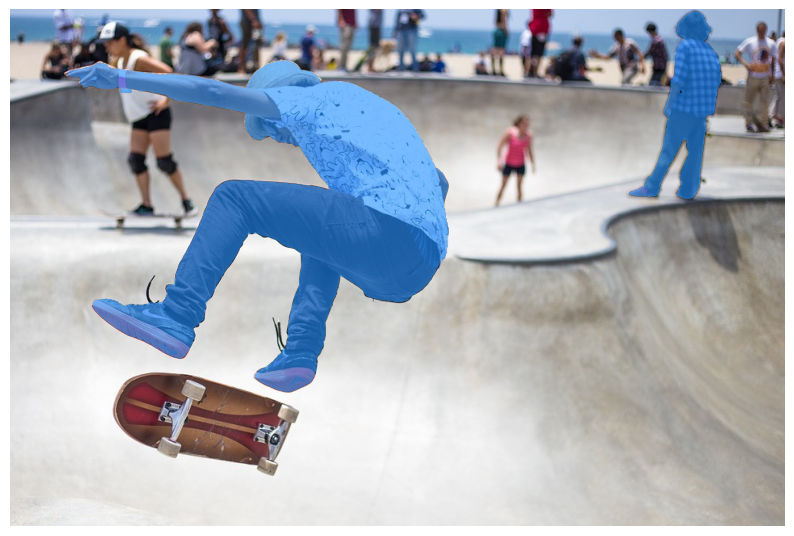} \\
        (a) original image & (b) instance segmentation \\
    \end{tabular}
    \caption{Example of a given image with areas of interest selected by bounding boxes and its corresponding masks after an instance segmentation. Adapted from Google Images.}
    \label{fig:sam_skate}
\end{figure}

In addition to this, another very interesting functionality for the purpose of this work is \textit{instance segmentation}, which uses points or bounding boxes to select an area of interest in the image and extract related masks from it, as demonstrated in Fig.~\ref{fig:sam_skate}. After selecting the bounding boxes, simply pass them as a parameter for the predict function to make the prediction.

\subsection{Model Development}

In the development of the model, three stages were used to create the masks that make up the database used during training. First, labels were manually and interactively placed on each image mask in a browser, and after that, the network was trained 6 times.

In the second stage, called semi-automatic, the masks were delivered to the involved professionals, who had the function of checking if any object was not considered in the mask production process by the network. Again, the masks were fed into the network, which was trained 5 more times.

In the final stage, the production of masks was carried out automatically without any human intervention. At this stage, 32x32 points were evenly distributed throughout the image to produce masks covering the entire data extension. After completing this process, the masks underwent filtering and post-processing. In the filtering, one step was to remove masks covering 95\% or more of the image, as it does not make sense to keep a mask that identifies the image as a whole. On the other hand, in post-processing, masks with an area smaller than 100 pixels were removed~\cite{sam}.

\begin{figure}[!htb]
    \centering
    \includegraphics[width=\textwidth]{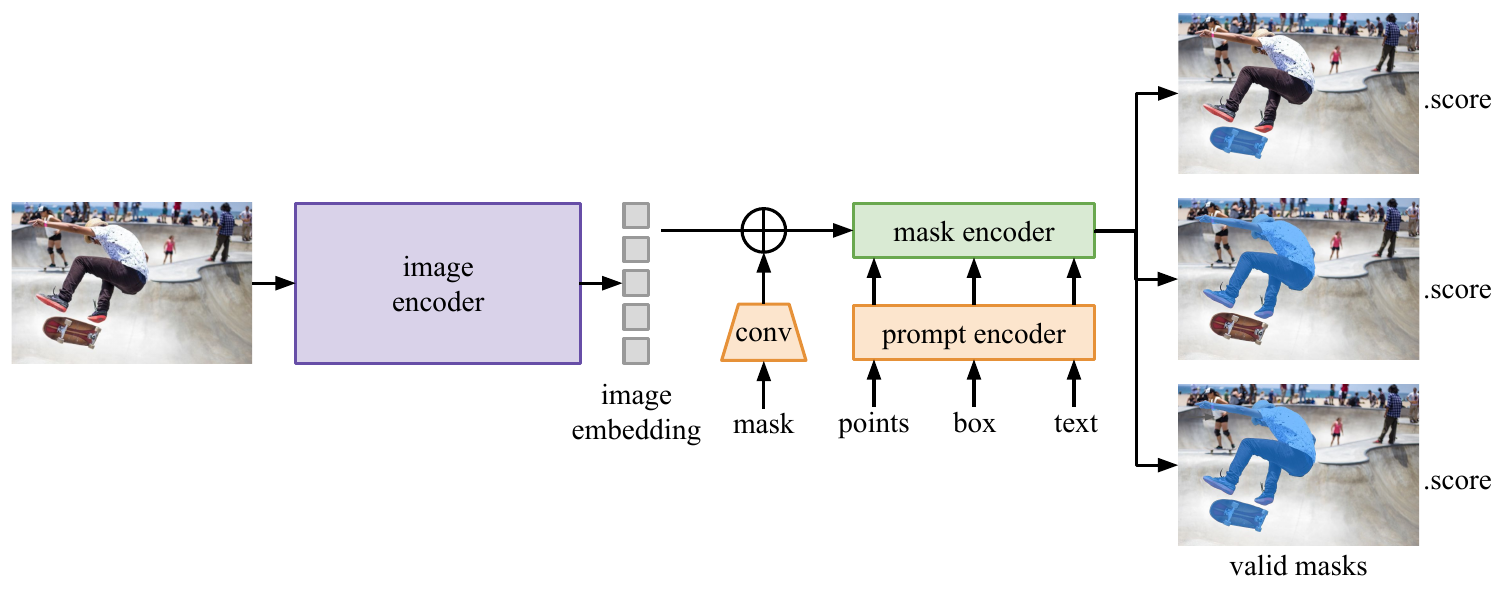}
    \caption{Organization of the Segment-Anything Model. Adapted from~\cite{sam}.}
    \label{fig:estrutura_sam}
\end{figure}

\subsection{Components of SAM}

SAM uses three components to perform mask prediction~\cite{sam}, as depicted in Fig.~\ref{fig:estrutura_sam}.

\subsubsection{Image Encoder}

This component is responsible for processing the input image. For this, a pre-trained Vision Transformer~(ViT) was used. The encoder is executed once per image and can be applied before the input is fed to decoder.

\subsubsection{Prompt Encoder}

This deals with the prompts provided to the model for the segmentation process. There are two sets of prompts considered: sparse (such as points and bounding boxes) and dense (masks). For representation purposes, sparse prompts use positional encodings and learned embeddings, while dense ones are processed by convolutions and summed with the image representation.

\subsubsection{Mask Decoder}

This component is responsible for generating the segmentation mask based on the encoded input information. This is done by modifying a transformer decoder block and layers that perform dynamic mask prediction.

\section{Materials and Methods}

\subsection{Datasets}

Since the procedure of identifying parts in an image will not be performed in this work, the proposal is to use publicly recognized datasets that have been reviewed by qualified medical professionals.

It is important to highlight the availability of actual masks provided by healthcare professionals in both datasets. The absence of these masks would hinder proper evaluation and comparison of the model's results, compromising the integrity of the analysis.

To work with these datasets, preprocessing of both the images and provided masks was necessary. Initially, to enable the use of SAM in predicting images, as well as fine-tuning and evaluating the model's accuracy, standardization of resolution to 256x256 was required. 

\subsubsection{Montgomery dataset~\cite{dados}}

Comprising 138 images, this dataset was acquired from the Maryland Department of Health and Human Services and published by the United States National Library of Medicine. It consists of chest X-ray images whose corresponding masks for the left and right lungs are stored separately and, for this reason, a pre-processing step was necessary to merge them.  

\subsubsection{Shenzhen dataset~\cite{dados}}

This dataset contains 566 chest X-ray images obtained from the Shenzhen Hospital, also published by the United States National Library of Medicine. 

\subsection{Technical Approach}

The approach used to improve SAM's performance, as previously mentioned, was fine-tuning. After conducting tests with different prompts for the adjustment, including bounding boxes, points extracted from average images, and a combination of both, it was found that the most effective approach was using sets of points obtained from the average images of each dataset. This is due to their better performance.

\begin{figure}
  \centering
  \includegraphics[width=0.38\textwidth]{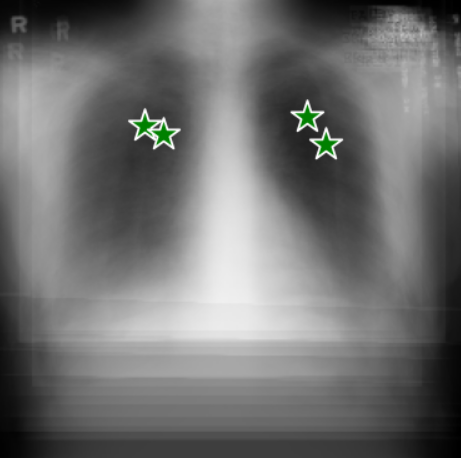}
  \caption{Example of a mean image obtained on the Montgomery dataset.}
  \label{fig:imagem_media}
\end{figure}

It is important to highlight that, to avoid any bias or interference in the validation and test sets, only the samples designated for training were used in constructing such images and points. This ensures an objective evaluation of the model's predictions. Fig \ref{fig:imagem_media} presents an example of a mean image obtained from the Montgomery dataset.

For the bounding boxes needed for training, a function was used to extract the boxes from the image masks and apply perturbations to their coordinates to improve the model's robustness and generalization.

During the fine-tuning process, it was necessary to ensure that gradients were calculated exclusively in the mask decoder, to avoid undesired changes in the other components of SAM. This approach aims to preserve the representations learned in the upper layers of the neural network, which may contain valuable information for the segmentation process.

\subsection{Evaluation Metrics}

\subsubsection{Intersection over Union~(IoU)}

In the Intersection over Union~(IoU) metric, the similarity between the mask produced by the model~($M_p$) and the provided ground truth mask~($M_v$) is evaluated. This is calculated as the ratio between the intersection and the union of the two masks:
\begin{equation}
\centering
\text{IoU}(M_v, M_p) = \frac{|M_v \cap M_p|}{|M_v \cup M_p|},
\end{equation}

\noindent where $M_v \cap M_p$ and $M_v \cup M_p$ are, respectively, the intersection and union between sets $M_v$ and $M_p$, and $| \cdot |$ is their cardinality.

\subsubsection{F1-Score}

The F1-Score, with a purpose similar to the previous metric, is an evaluation measure that combines precision and recall into a single value, given by the following expression:
\begin{equation}
\centering
\text{F1-Score}(M_v, M_p) = \frac{2 \cdot |M_v \cap M_p|}{|M_v| + |M_p|}.
\end{equation}

\subsection{Training Procedure}
\label{sec:training_procedure}

Before starting the experiments, the datasets were divided into validation, training, and test sets. This division is a common practice in machine learning to evaluate and adjust the training and applied techniques. In this work, 20\% of the data was set aside for testing, 60\% for training, and 20\% for validation. The validation set was used to adjust some hyperparameters during the process, such as the segmentation threshold and the loss function.

For comparison with the results obtained with the U-Net network~\cite{unet}, a 5-fold cross-validation was performed in the same manner as the work of Brioso~\cite{trab_comparacao}. Cross-validation is a technique to assess the generalization capability of the model on a given dataset. In short, cross-validation splits the data into several subsets called folds. After this procedure, the model is trained, tested, and validated with different portions of the data until all parts have been used as test sets. At the end of the process, the evaluation metrics, such as F1-Score and IoU, are aggregated from all the training and testing iterations to provide a more robust and less biased estimate of performance.

\subsection{Loss Function and Optimizer}

Regarding the loss function, the ``\textit{DiceFocalLoss}'' from the Monai library\footnote{\url{https://monai.io/} (As of July, 2024)} was chosen, which computes both Dice and Focal losses and returns a weighted sum of them. This approach demonstrated superior performance compared to other available loss functions for image segmentation, such as ``\textit{DiceCELoss}'', which combines Dice and Cross Entropy losses. 
As for the optimizer, Adam was chosen, a widely-used method that combines the benefits of RMSProp and SGD Momentum to converge quickly to the global minimum.

\section{Experiments and Results}
\label{sec:experimentos}

This section presents the experiments conducted to evaluate and optimize SAM's performance. Initially, the results of the model before fine-tuning are discussed, followed by an analysis of the learning curve. Subsequently, we explore the influence of the segmentation threshold on the quality of the masks generated by SAM and detail the hyperparameter tuning process that optimized the model's performance. Finally, experiments about the model's adaptability and its comparison with other baselines are discussed. 

\subsection{Initial Evaluation of SAM}

Before fine-tuning SAM for lung segmentation in chest X-ray images, an initial evaluation was conducted to measure the performance of the neural network in its original form~\cite{sam}. This step is important to establish a baseline reference for the model with respect to the Montgomery and Shenzhen datasets~\cite{dados}.
For this, the datasets were divided into five subsets to follow the 5-fold cross-validation procedure. Tables~\ref{tab:avaliacao_inicial_mont}~and~\ref{tab:avaliacao_inicial_shen} present the mean and standard deviation of the evaluation metrics calculated on the resulting subsets obtained from the Montgomery and Shenzhen datasets, respectively.

\begin{table}[!htb]
	\centering
	\caption{Initial evaluation of SAM on the Montgomery dataset.}
	\label{tab:avaliacao_inicial_mont}
	\begin{tabular}{|c|c|c|c|}
		\hline 
		\textbf{Prompt} & \textbf{F1-Score} & \textbf{IoU} \\
		\hline
		\hline \textbf{Bounding Box} & $0.718\pm 0.033$ & $0.586\pm 0.033$ \\
		\hline \textbf{Points} & $0.860\pm 0.013$ & $0.774\pm 0.018$ \\
		\hline \textbf{Both} & $0.848\pm 0.006$ & $0.746\pm 0.007$ \\
		\hline
	\end{tabular}
\end{table}

\begin{table}[!htb]
    \centering
    \caption{Initial evaluation of SAM on the Shenzhen dataset.}
    \label{tab:avaliacao_inicial_shen}
	\begin{tabular}{|c|c|c|c|}
		\hline 
		\textbf{Prompt} & \textbf{F1-Score} & \textbf{IoU} \\
		\hline
		\hline \textbf{Bounding Box} & $0.782\pm 0.009$ & $0.661\pm 0.013$ \\
		\hline \textbf{Points} & $0.726\pm 0.021$ & $0.593\pm 0.026$ \\
		\hline \textbf{Both} & $0.863\pm 0.005$ & $0.765\pm 0.008$ \\
		\hline
	\end{tabular}
\end{table}

\subsection{Learning Curve}

To achieve satisfactory results, various experiments were conducted during the model's training procedure. Initially, to determine the appropriate number of epochs for fine-tuning SAM, the learning curve was used. This curve represents the average loss obtained by the neural network over the training epochs and is an essential tool for evaluating its progress and performance. The learning curve allows analysis, through the decrease in loss over time, of whether the model is optimizing over the epochs or stagnating, the latter indicating that the training can be stopped.

\begin{figure}
  \centering
  \includegraphics[width=0.48\textwidth]{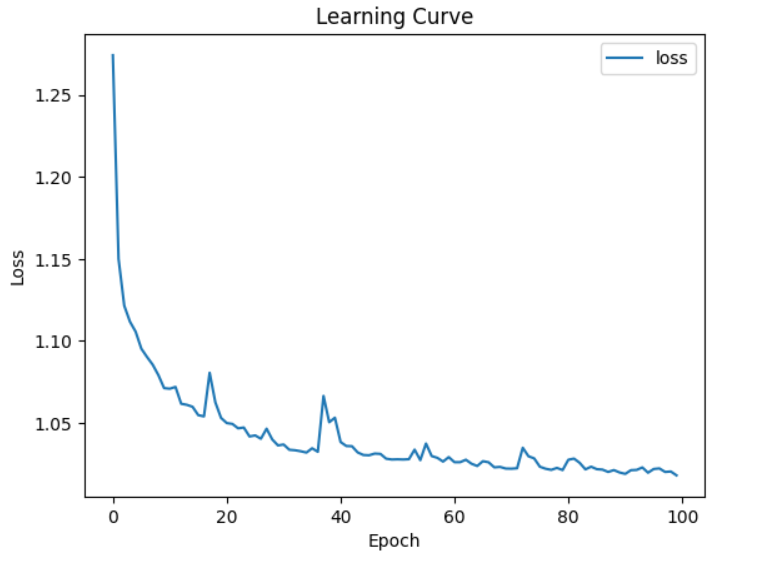}
  \caption{Example of a learning curve obtained on the Shenzhen dataset.}
  \label{fig:curva_aprendizado}
\end{figure}

From the aforesaid tests, it was found that none of the datasets and none of the different prompts evaluated showed a significant improvement in loss after the hundredth epoch. Fig~\ref{fig:curva_aprendizado} shows an example of one of the learning curves obtained on the Shenzhen dataset.

\subsection{Segmentation Threshold}

Another critical factor influencing SAM's performance after fine-tuning is the threshold used to convert the segmented mask into a binary mask after model's prediction. This is because the mask returned by the model has continuous values between 0 and 1, representing the probability of a given pixel belonging to the region of interest. The choice of threshold to binarize this mask directly affects the results, as an inappropriate selection of this value can lead to inaccurate or incomplete segmentation masks. Low values (e.g., below 0.5) may include many pixels that do not belong to the mask, while high values (e.g., above 0.7) may exclude many pixels that belong to the mask. For this reason, in this study, threshold values from 0.50 to 0.70 were evaluated on the validation set.

\begin{table}[!htb]
    \centering
    \caption{F1-Score obtained for different thresholds in the Montgomery dataset.}
    \label{tab:limiar_mont}
    \begin{tabular}{|c|c|c|c|c|}
        \hline \textbf{Threshold} & \textbf{Bounding Box} & \textbf{Points} & \textbf{Both} \\
        \hline
        \hline \textbf{0.50} & \textbf{0.828} & 0.898 & \textbf{0.894} \\
        \hline \textbf{0.55} & 0.812 & 0.932 & 0.893 \\
        \hline \textbf{0.60} & 0.743 & 0.957 & 0.879 \\
        \hline \textbf{0.65} & 0.596 & \textbf{0.960} & 0.835 \\
        \hline \textbf{0.70} & 0.410 & 0.938 & 0.759\\
        \hline
    \end{tabular}
\end{table}

The results obtained for the Montgomery and Shenzhen datasets are presented in Tables~\ref{tab:limiar_mont}~and~\ref{tab:limiar_shen}, respectively. 
The best F1-Score for each dataset and prompt is highlighted in bold. 
This information helped identify the most suitable threshold value for each dataset and input, thereby improving the overall accuracy and quality of predictions.

\begin{table}[!htb]
    \centering
    \caption{F1-Score obtained for different thresholds in the Shenzhen dataset.}
    \label{tab:limiar_shen}
    \begin{tabular}{|c|c|c|c|c|}
        \hline \textbf{Threshold} & \textbf{Bounding Box} & \textbf{Points} & \textbf{Both} \\
        \hline
        \hline \textbf{0.50} & \textbf{0.837} & 0.880 & \textbf{0.846} \\
        \hline \textbf{0.55} & 0.127 & \textbf{0.918} & 0.782 \\
        \hline \textbf{0.60} & 0.003 & 0.870 & 0.417 \\
        \hline \textbf{0.65} & 0.000 & 0.829 & 0.105 \\
        \hline \textbf{0.70} & 0.000 & 0.789 & 0.005 \\
        \hline
    \end{tabular}
\end{table}

\subsection{Hyperparameter Tuning}

Hyperparameter tuning is a crucial step in the SAM fine-tuning process as it determines the values of learning rate and weight decay, which directly impact the model's performance. 
Briefly, the learning rate controls the size of steps that the optimization algorithm takes during the learning process. A too high value can lead to instability and prevent or hinder algorithm convergence, whereas a too low value can result in slow or stagnant training. On the other hand, weight decay controls the magnitude of the model's weights by adding a penalty to prevent them from becoming too large, which can help prevent overfitting.

In this study, a grid of predefined values was used: [$1 \times 10^{-5}$, $1 \times 10^{-4}$, $1 \times 10^{-3}$] for the learning rate and [0, $1 \times 10^{-1}$, $1 \times 10^{-3}$] for the weight decay.
After the search, a learning rate of $1 \times 10^{-5}$ and a weight decay of 0 were adopted.

\subsection{Model Adaptability}

To verify the model's adaptability, two experiments were conducted in which SAM was trained on one dataset and tested on another. This procedure allows for evaluating the model's generalization capability in different contexts, which is important for ensuring utility and effectiveness in real-world situations where data may vary. It is worth noting that, unlike the other tests, this was performed only with the points obtained from the average image.

First, fine-tuning was performed on the Shenzhen dataset and then applied to the Montgomery dataset. The results of this experiment can be seen in Table~\ref{tab:adapt_shen}. For a comparison of the results, the F1-Score and IoU values for the Montgomery dataset trained with its own images were also included.

\begin{table}[!htb]
    \centering
    \caption{Adaptability of the model trained on the Shenzhen dataset and tested on the Montgomery dataset.}
    \label{tab:adapt_shen}
    \begin{tabular}{|c|c|c|}
        \hline \textbf{Metric} & \textbf{Shenzhen $\rightarrow$ Montgomery} & \textbf{Montgomery $\rightarrow$ Montgomery} \\
        \hline
        \hline \textbf{F1-Score} & 0.924 & 0.943 \\
        \hline \textbf{IoU} & 0.860 & 0.897 \\
        \hline
    \end{tabular}
\end{table}

Similarly, another experiment was conducted where the neural network was initially fine-tuned on the Montgomery dataset and then applied to the Shenzhen dataset. Interestingly, the evaluated results were better compared to the fine-tuning performed directly on the Shenzhen dataset.

\begin{table}[!htb]
    \centering
    \caption{Adaptability of the model trained on the Montgomery dataset and tested on the Shenzhen dataset.}
    \label{tab:adapt_mont}
    \begin{tabular}{|c|c|c|}
        \hline \textbf{Metric} & \textbf{Montgomery $\rightarrow$ Shenzhen} & \textbf{Shenzhen $\rightarrow$ Shenzhen} \\
        \hline
        \hline \textbf{F1-Score} & 0.933 & 0.915 \\
        \hline \textbf{IoU} & 0.875 & 0.845 \\
        \hline
    \end{tabular}
\end{table}

\subsection{Best Models}

Tables~\ref{tab:result_mont}~and~\ref{tab:result_shen} present the results of the best models for the Montgomery and Shenzhen datasets, respectively. 
They were trained using the best hyperparameters found in the previously conducted experiments and with different prompts.
The results were obtained based on 5-fold cross-validation, as described in Section~\ref{sec:training_procedure}. Additionally, to measure the variability of the results with respect to the mean, the standard deviation is also indicated.

\begin{table}[!htb]
    \centering
    \caption{Detailed results on the Montgomery dataset.}
    \label{tab:result_mont}
    \begin{tabular}{|c|c|c|c|}
        \hline \textbf{Prompt} & \textbf{F1-Score} & \textbf{IoU} \\
        \hline
        \hline \textbf{Bounding Box} & $0.818\pm 0.040$ & $0.707\pm 0.045$ \\
        \hline \textbf{Points} & $0.943\pm 0.007$ & $0.897\pm 0.012$ \\
        \hline \textbf{Both} & $0.876\pm 0.015$ & $0.787\pm 0.023$ \\
        \hline
    \end{tabular}
\end{table}

\begin{table}[!htb]
    \centering
    \caption{Detailed results on the Shenzhen dataset.}
    \label{tab:result_shen}
    \begin{tabular}{|c|c|c|c|}
        \hline \textbf{Prompt} & \textbf{F1-Score} & \textbf{IoU} \\
        \hline
        \hline \textbf{Bounding Box} & $0.797\pm 0.014$ & $0.667\pm 0.024$ \\
        \hline \textbf{Points} & $0.915\pm 0.011$ & $0.845\pm 0.018$ \\
        \hline \textbf{Both} & $0.845\pm 0.012$ & $0.735\pm 0.018$ \\
        \hline
    \end{tabular}
\end{table}

\subsection{Comparison with U-Net}

A comparison with the best results reported by Brioso~\cite{trab_comparacao} for by U-Net network~\cite{unet} is also performed.
The values used for comparison with U-Net represent the best performance achieved by SAM on the test set. These results correspond to the models trained with points extracted from the average images.

\begin{table}[!htb]
    \centering
    \caption{SAM compared to U-Net (F1-Score).}
    \label{tab:comp_dice_coef}
        \begin{tabular}{|c|c|c|}
           \hline 
            \textbf{Dataset} & \textbf{SAM} & \textbf{U-Net} \cite{trab_comparacao} \\
            \hline
            \hline \textbf{Montgomery} & $0.943\pm 0.007$ & \textbf{$0.973\pm 0.014$} \\
            \hline \textbf{Shenzhen} & $0.915\pm 0.011$ & \textbf{$0.941\pm 0.047$} \\
            \hline
        \end{tabular}
\end{table}

Above, in Table~\ref{tab:comp_dice_coef}, F1-Score metric values are compared between SAM and U-Net, as presented in~\cite{trab_comparacao}. The results indicate that the U-Net network, known for its effectiveness in this task, achieved superior performance on both the Montgomery and Shenzhen datasets.
Unlike the findings of He~et~al.~\cite{acuracia_sam}, despite a slightly lower F1-Score, SAM demonstrated satisfactory results similar to those of U-Net, indicating its feasibility and potential for future applications.

\subsection{Predicted Masks}

\begin{figure}
  \centering
  \includegraphics[width=\textwidth]{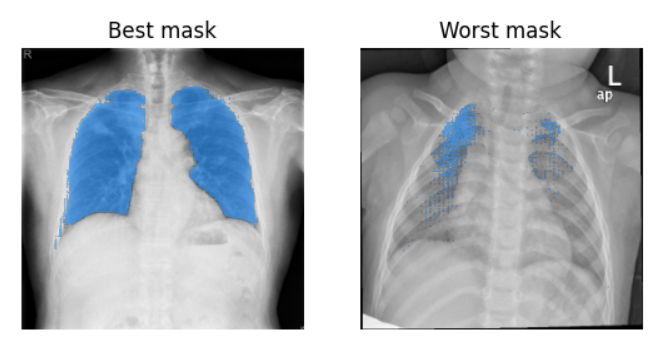}
  \caption{Best and worst masks obtained on the Shenzhen dataset, respectively.}
  \label{fig:melhor_pior_shen}
\end{figure}

\begin{figure}
  \centering
  \includegraphics[width=\textwidth]{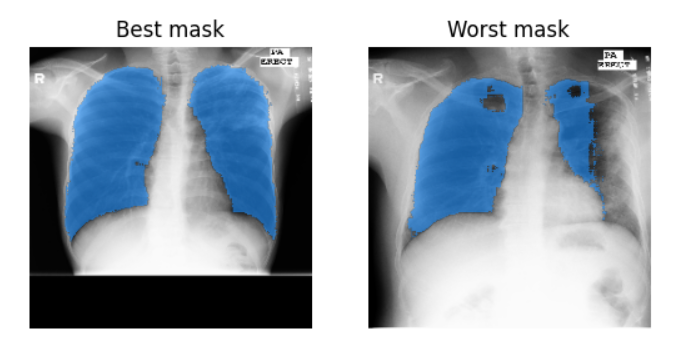}
  \caption{Best and worst masks obtained on the Montgomery dataset, respectively.}
  \label{fig:melhor_pior_mont}
\end{figure}

Finally, for illustration purposes, Fig.~\ref{fig:melhor_pior_shen} shows the best and worst predictions performed by SAM on the Shenzhen dataset.
Similarly, in Fig.~\ref{fig:melhor_pior_mont}, the best and worst predictions obtained on the Montgomery dataset can be seen.

\section{Conclusions}

In light of the promising neural network proposed by Meta AI, this work sought ways to fine-tune SAM and adapt it for lung segmentation in chest X-ray images provided by the Shenzhen and Montgomery datasets~\cite{dados}. Its main objective is to contribute to the automation and accuracy of medical diagnosis, as well as to provide insights and opportunities regarding this new technology.

Throughout this study, various strategies to fine-tuning SAM were investigated, including the use of different prompts, such as bounding boxes, points selected from the average image, and a combination of both. Additionally, different numbers of training epochs and loss functions were explored, with the latter being a crucial choice for satisfactory results.

The experiments conducted during the mask prediction phase allowed for evaluating the impact of the threshold used in the binarization process, that is, converting the masks from the soft mask format to the hard mask format. This analysis provided valuable information on the model's sensitivity to different thresholds and their significant influence on mask quality. Furthermore, tests conducted to verify the model's adaptability showed its utility in real-world situations where datasets exhibit variations.

Considering the complexity of the challenging task of segmenting lungs in chest X-ray images, future work can explore different transfer learning techniques not covered in this study to further improve the model’s performance.

\begin{credits}
\iffinal
\subsubsection{\ackname} This research was supported by São Paulo Research Foundation - FAPESP (grant 2023/17577-0) and Brazilian National Council for Scientific and Technological Development - CNPq (grants 315220/2023-6, 420442/2023-5, and 146570/2023-5).
\fi

\subsubsection{\discintname}
The authors have no competing interests to declare that are relevant to the content of this work.
\end{credits}

%
%
%
\bibliographystyle{splncs04}
%

\end{document}